\documentstyle[aaspp4,flushrt]{article}
\input psfig.sty

\makeatletter

\@ifundefined{chapter}{\def\thebibliography#1{\section*{References\@mkboth
  {REFERENCES}{REFERENCES}}\list
  {\relax}{\setlength{\labelsep}{0em}
        \setlength{\itemindent}{-\bibhang}
        \setlength{\itemsep}{0pt}
        \setlength{\parsep}{0pt}
        \setlength{\leftmargin}{\bibhang}}
    \def\newblock{\hskip .11em plus .33em minus .07em}
    \sloppy\clubpenalty4000\widowpenalty4000
    \sfcode`\.=1000\relax}}%
{\def\thebibliography#1{\chapter*{Bibliography\@mkboth
  {BIBLIOGRAPHY}{BIBLIOGRAPHY}}\list
  {\relax}{\setlength{\labelsep}{0em}
        \setlength{\itemindent}{-\bibhang}
        \setlength{\itemsep}{0pt}
        \setlength{\parsep}{0pt}
        \setlength{\leftmargin}{\bibhang}}
    \def\newblock{\hskip .11em plus .33em minus .07em}
    \sloppy\clubpenalty4000\widowpenalty4000
    \sfcode`\.=1000\relax}}

\newlength{\bibhang}
\let\@internalcite\cite
\def\cite{\let\@citeleft(\let\@citeright)%
    \@ifstar{\citeyear}{\citefull}}
\def\citenp{\let\@citeleft\relax\let\@citeright\relax
    \@ifstar{\citeyear}{\citefull}}
\def\citefull{\def\astroncite##1##2{##1~##2}\@internalcite}
\def\citeyear{\def\astroncite##1##2{##2}\@internalcite}

\def\@citex[#1]#2{\if@filesw\immediate\write\@auxout{\string\citation{#2}}\fi
  \def\@citea{}\@cite{\@for\@citeb:=#2\do
    {\@citea\def\@citea{; }\@ifundefined
       {b@\@citeb}{{\bf ?}\@warning
       {Citation `\@citeb' on page \thepage \space undefined}}%
{\csname b@\@citeb\endcsname}}}{#1}}

\def\@cite#1#2{\@citeleft#1\if@tempswa , #2\fi\@citeright}
\def\@biblabel#1{}

\makeatother

\def\gsim{\;\rlap{\lower 2.5pt
 \hbox{$\sim$}}\raise 1.5pt\hbox{$>$}\;}
\def\lsim{\;\rlap{\lower 2.5pt
   \hbox{$\sim$}}\raise 1.5pt\hbox{$<$}\;}

\def\angle0{\mbox{\boldmath$\theta_{\rm planet}$}}

\def\Deltabeta{\mbox{\boldmath$\Deltabeta$}}

\begin{document}

\title{The Cosmic Baryon Fraction and the Extragalactic Ionizing Background}
\author{Lam Hui$^{a}$, Zolt\'an Haiman$^{b}$, Matias Zaldarriaga$^{c}$ and Tal Alexander$^{d}$
\\
\vspace{0.2cm}
$^{a}$ Department of Physics, Columbia University, 538 West 120th Street,
New York, NY 10027\\   
$^{b}$ Princeton University Observatory, Princeton, NJ 08544\\
$^{c}$ Physics Department, New York University, 4 Washington Place, New York, NY 10003\\
$^{d}$ Space Telescope Science Institute, 3700 San Martin Drive, Baltimore, MD 21218\\
{\tt lhui@astro.columbia.edu, zoltan@astro.princeton.edu, matiasz@physics.nyu.edu, tal@stsci.edu}
}

\begin{abstract}
We reassess constraints on the cosmological baryon density from observations of
the mean decrement and power spectrum of the Lyman-$\alpha$ forest, taking into
account uncertainties in all free parameters in the simplest 
gravitational instability
model.  The uncertainty is dominated by that of the photoionizing background,
but incomplete knowledge of the thermal state of the intergalactic medium also
contributes significantly to the error-budget.  While 
current estimates of the baryon fraction from the forest
do prefer values that are somewhat higher than the big bang
nucleosynthesis value of $\Omega_b h^2 = 0.02 \pm 0.001$,
the discrepancy is at best about $3 \sigma$. For instance, assuming the 
highest estimate of the ionizing background, 
as indicated by recent measurements of a large escape fraction from 
Lyman-break galaxies by Steidel, Pettini \& Adelberger, we find 
$\Omega_b h^2 = 0.045 \pm 0.008$. 
A recent measurement of the ionizing background from the proximity effect
by Scott et al., on the other hand, implies $\Omega_b h^2 = 0.03 \pm 0.01$. 
We provide an expression from which future likelihoods for
$\Omega_b h^2$ can be derived as measurements of the ionizing
background improve -- consistency
among constraints from the forest, nucleosynthesis and the 
microwave background will provide a powerful test of the
gravitational instability model for the forest, and for
large scale structure in general.
We also develop a formalism
which treats lower bounds on the baryon density in a statistical manner, which
is appropriate if only a lower bound on the ionizing background is
known. Finally, we discuss the implications of the escape fraction measurement
for the age, structure and stellar content of Lyman-break galaxies.
\end{abstract}

\section{Introduction}
\label{intro}

It has long been recognized that the intergalactic medium (IGM) is highly
ionized, and likely contains a substantial fraction of the universe's baryons
at redshifts $z\sim 3$ (e.g. Gunn \& Peterson 1965, Steidel \& 
Sargent 1987, Fukugita, Hogan \& Peebles
1998 and references therein). Constraints on the baryon density $\Omega_b h^2$
(where $\Omega_b$ is a fraction of the critical density, and $h$ is the Hubble
constant today in units of $100$ km/s/Mpc) from observations of the 
Lyman-$\alpha$
(Ly$\alpha$) forest rely on assumptions about the size and geometry of the
absorbing structures (see e.g.  Rauch \& Haehnelt 1995). Recent advance in
theoretical modeling of the forest provides a framework in which fluctuations
seen in absorption arise naturally out of gravitational instability, and which
successfully matches the gross observed properties of the forest
(e.g.
\citenp{bi92,cen94,zhang95,rm95,hernquist96,jordi96,muecket96,bi97,bond97,croft98,hui97a,theuns99,bryan99,croft99,nusser00}; but see also
Meiksin, Bryan \& Machacek 2001 for possible problems, and 
Zaldarriaga et al. 2001 for consistency checks
). 
This framework makes definite predictions about the spatial distribution of the
absorbing structures, making it possible to obtain useful constraints on
$\Omega_b h^2$ from forest observations
(e.g. \citenp{rauch97,weinberg97,mcdonald99}) -- once the photoionizing
background, which determines the overall neutral fraction, is specified.

Three recent developments motivate us to re-examine the Ly$\alpha$ forest
constraints on the cosmological baryon fraction.  First, the recent microwave
background anisotropy measurements favor a baryon fraction that is higher than
the big bang nucleosynthesis value inferred from deuterium measurements (the
former gives $0.022 < \Omega_b h^2 < 0.040$, 2-$\sigma$ limits for a flat
scale-invariant inflation model, see
\citenp{boom00,maxima00,tz00,jaffe00,lange01}; 
the latter gives $0.02 \pm 0.001$, 1-$\sigma$ error, see Burles, Nollet \&
Turner 2000; see also Burles \& Tytler 1998 for earlier
constraints).
Second, Steidel, Pettini \& Adelberger
(2001) reported measurements of a large escape fraction of ionizing photons
from Lyman break galaxies (LBG's) at $z \sim 3$, which imply an extragalactic
ionizing flux that is significantly larger than what one might expect from
quasars alone. A larger photo-ionizing background implies a smaller neutral
fraction, and so a larger baryon density is required to match the given
observed mean absorption seen in the forest.  
Lastly, analysis of the Ly$\alpha$ forest has moved beyond comparisons with a
restricted set of cosmological models
to a point where several parameters specifying the mass
power spectrum and the thermal state of the IGM can be determined at the same
time from the data (e.g.  Choudhury, Srianand \& Padmanabhan 2000, Zaldarriaga,
Hui \& Tegmark 2000). Since earlier forest constraints on the baryon fraction
focused on fixed cosmological models and/or reionization history (and therefore
fixed thermal state)
\footnote{A notable exception is Weinberg et al. (1997) who attempted to give
the most conservative lower bounds on the baryon density.  We provide here a
statistical framework for interpreting such lower bounds.}, it is timely to
examine how uncertainties in these parameters impact the estimates for
$\Omega_b h^2$. Because the Ly$\alpha$ constraint on the baryon fraction
is completely independent of constraints from nucleosynthesis or
the microwave background, comparing all three provides a powerful
consistency check of the inflation $+$ cold dark matter structure
formation model, as well as a test of current ideas about
the nature of the forest.

We briefly review in \S \ref{model} the gravitational instability model for the
Ly$\alpha$ forest, and how it is used to obtain estimates of the baryon
density. Discussions in the literature have focused on two different
approaches. 
One (e.g. Haehnelt et al. 2000)
implicitly assumes that most of the universe's baryons is in the IGM
\footnote{Estimates so far do give values for the baryon density that are close
to or even higher the the nucleosynthesis value (e.g. Rauch et al. 1997,
McDonald et al 2000). Detailed accounting of the universe's baryon budget also
supports the notion that most of the baryons are in the IGM (Fukugita et
al. 1998).}, and that the total ionizing background can be largely accounted
for by directly summing up the contributions from known sources, or obtained
from indirect measurements of the sum total background, such as those using the
proximity effect.
The other approach (e.g. Rauch et al. 1997; Weinberg et al. 1997)
is more conservative, assuming only a lower limit to the
ionizing background from known sources -- this gives a lower bound on the
neutral fraction and therefore a lower bound for the $\Omega_b h^2$.  We will
discuss both approaches.

We describe in \S \ref{lowerbound} how lower bounds given in the literature can
be interpreted in a statistical manner, generalizing earlier work where only
strict lower bounds are discussed.  As we will see, uncertainties in the level
of the ionizing background dominate uncertainties in the current bound on
$\Omega_b h^2$.  In \S \ref{further}, we relate the lower-bound likelihood in
\S \ref{lowerbound} to the (differential) likelihood for baryon density,
assuming effectively that most of the universe's baryons are in the IGM and
that the ionization background is known.  As we will emphasize in \S
\ref{further}, non-negligible uncertainties in the estimates for $\Omega_b h^2$
remain {\it even if} the ionizing background is known to high accuracy. This is
because of degeneracy with other parameters in the gravitational instability
model. We will examine which parameters are the most important in this
regard. The likelihood of $\Omega_b h^2$ for a fixed ionizing background is
also given (an explicit expression is also presented in \S \ref{discuss}),
which can be used to compute future likelihoods as measurements of the ionizing
background improve.

Some of the conclusions in the above sections depend critically on the ionizing
flux inferred from observations of Lyman break galaxies (Steidel et
al. 2001). The fraction of hydrogen--ionizing photons that escape from these
galaxies is significantly higher than what is observed in local galaxies, and
what is naively expected in simple theoretical models.  In \S \ref{LBO}, we
investigate the type of cold gas distribution that would
allow these early galaxies to have a high escape probability for ionizing photons.
Finally, we conclude in \S \ref{discuss}, where the issue of escape fraction
from quasars is also discussed.

Some of the issues in this paper have been addressed in two recent works.
Haehnelt et al. (2000) used previous bounds to derive new limits on $\Omega_b
h^2$, using the Steidel et al. measurement of the escape fraction -- the
treatment assumed the same cosmological model and reionization history upon
which the previous respective bounds were based.  They also briefly discussed
the significance the Steidel et al. results in the context of population
synthesis model spectra, and the escape fraction expected based on the amount
of gas inferred from the star--formation rate. Choudhury, Srianand \&
Padmanabhan (2001) used the lognormal approximation to derive bounds on
the ratio of $(\Omega_b h^2)^2$ to the ionizing flux, where the thermal
state of the gas is allowed to vary, for a fixed cosmological model.
Here, we examine the $\Omega_b h^2$
constraints using N-body simulations for a suite of cosmological models
(i.e. varying power spectrum) and thermal histories.  We also consider the
implication of the high observed escape fraction for the age, structure and
stellar content of Lyman-break galaxies in greater detail.

\section{Baryon Fraction from the Lyman {\protect\bf $\alpha$} Forest}
\label{baryon}

\subsection{The Gravitational Instability Model for the Forest}
\label{model}

We describe briefly here the Ly$\alpha$ forest model
used in this paper.
It is largely motivated by the success of numerical simulations in
matching the observed properties of quasar absorption spectra
(see Cen et al. 1994 and related references in \S \ref{intro}).
The Ly$\alpha$ optical depth $\tau$ along the line
of sight is given by $A (1+\delta)^\alpha$, mapped to redshift space,
and 
smoothed with the thermal broadening window
(see e.g. Hui, Gnedin \& Zhang 1997 for details). The quantity
$\delta$ is the baryon overdensity $\delta \rho/\bar\rho$. 
The constant $\alpha$ is given by $2- 0.7 (\gamma -1)$ where
$\gamma$ is the equation of state index -- i.e.
the temperature of the gas $T$ is related to density by
$T = T_0 (1+\delta)^{\gamma - 1}$, 
where $T_0$ is the temperature at mean density (Croft et al. 1997, Hui \& Gnedin 1997).
The constant $A$ is given by (e.g. Weinberg et al. 1997)
\footnote{$A = \bar n_{\rm HI} c H(z)^{-1} \sigma_{\alpha 0}$,
where $\bar n_{\rm HI}$ is the neutral hydrogen density
at $\delta=0$, $c$ is the speed of light,
and $H(z)$ is the Hubble parameter at redshift $z$,
and $\sigma_{\alpha 0} =
4.478 \times 10^{-18} {\,\rm cm^2}$: 
the Ly$\alpha$ absorption cross-section is given by 
$\sigma_{\alpha 0}$ times the Voigt profile.
}
\begin{eqnarray}
\label{A}
A = && 0.18
\left[{\Omega_b h^2 \over 0.02}\right]^2 
\left[1 \over J_{\rm HI}\right]
\left[{T_0 \over 10^4 {\, \rm K}}\right]^{-0.7}
\left[{0.7 \over h}\right]
\left[{1+z \over 4}\right]^6 
\\ \nonumber &&
\left[X_{\rm H} \over 0.76\right]
\left[{X_{\rm H} + X_{\rm He}/2 \over 0.88}\right]
\left[{4.461 \over \sqrt{\Omega_m (1+z)^3 + \Omega_\Lambda
+ (1-\Omega_m -\Omega_\Lambda) (1+z)^2} }\right]
\end{eqnarray}
where $h$ is the Hubble constant today in units of $100$ km/s/Mpc,
$z$ is the redshift, $X_{\rm H}$ is the mass fraction of baryons
that is in hydrogen, $X_{\rm He}$ is the same for helium, the
assumption being that both H and He are highly
ionized.\footnote{Assuming instead that He is only singly ionized,
which amounts to changing $2 X_{\rm He}$ to $X_{\rm He}$, will make
little difference given our other sources of uncertainties.
Whether helium is singly ionized or not of course has profound
implications for the spectrum of the ionizing radiation -- that
is taken into account by the parameter $J_{\rm HI}$.
}
The factor of $4.461$ in
the last term corresponds to a model with $\Omega_m = 0.3$, $\Omega_\Lambda
=0.7$ and $z = 3$. We have used the recombination rate coefficient $\alpha_{\rm
rec.} = 4.2 \times 10^{-13} (T/10^4 {\, K})^{-0.7} {\, \rm cm^3 \, s^{-1}}$
(Rauch et al. 1997).

The dimensionless quantity $J_{\rm HI}$ is defined by
\begin{eqnarray}
\label{JHI}
&& J_{\rm HI} \times 10^{-21} {\, \rm erg \, s^{-1} \, cm^{-2} \, Hz^{-1} \,
ster^{-1}} \\ \nonumber 
= &&
\int_{\nu_{\rm HI}}^\infty 4 \pi j_\nu \sigma_{\rm HI} (\nu) 
{d\nu \over \nu} / 
\int_{\nu_{\rm HI}}^\infty 4 \pi \sigma_{\rm HI} (\nu) {d\nu \over \nu}
= 
j_{\nu_{\rm HI}} \left[ 3 \over \beta + 3 \right]
\end{eqnarray}
where $j_\nu$ is the specific intensity of the ionizing
background, $\nu_{\rm HI} = c/912 \AA$,
$\sigma_{\rm HI}$ is the ionizing cross-section for HI
($\sigma_{\rm HI} (\nu)$ is approximately 
$6.3 \times 10^{-18} {\,\rm cm^2 \,}
(\nu/\nu_{\rm HI})^{-3}$ for $\nu$ just above $\nu_{\rm HI}$; but see below), 
and $\beta$ is the slope of $j_\nu$ just above $\nu_{\rm HI}$:
$j_{\nu} \propto \nu^{-\beta}$. 
The photoionization rate $\Gamma_{\rm HI}$ is related to
$J_{\rm HI}$ by 
\begin{equation}
\Gamma_{\rm HI} = 4.3 \times 10^{-12} J_{\rm HI} {\, \rm s^{-1}}
\label{Gamma}
\end{equation}
The expression here, together with eq. (\ref{JHI}) above,
gives the correct relationship between $j_{\nu_{\rm HI}}$ 
and $\Gamma_{\rm HI}$ that is accurate to within $3 \%$, for
all values of $\beta$ between $0$ and $3$. This takes into
account slight departure of $\sigma_{\rm HI} (\nu)$ from
an exact $\nu^{-3}$ power-law (see Osterbrock 1989, eq. 2.4).

Given a structure formation model such as the Cold Dark Matter (CDM) model,
the distribution of $\delta$ follows from gravitational
instability, and can be predicted using
hydrodynamic, N-body (with suitable smoothing) or 
Hydro-PM (Gnedin \& Hui 1998) simulations.
Combining the $\delta$ fluctuations with the 
relation for the optical depth $\tau = A (1 + \delta)^\alpha$
(with suitable mapping to redshift space and thermal broadening),
one can predict a host of observable properties for the Ly$\alpha$ forest, 
for any given value of $A$ and $\alpha$.
Two observables that have attracted a lot of attention and have been
well measured are the transmission power spectrum and the mean decrement
($1- \langle e^{-\tau} \rangle$).
These measurements can be used to yield constrains on parameters in
the model, including $A$ which is proportional to $(\Omega_b h^2)^2$.
 
In this paper, we make use of the analysis by 
Zaldarriaga, Hui \& Tegmark (2000), 
who obtained constraints on 6 different parameters by matching
the observed transmission power spectrum and mean decrement at $z=3$ from
McDonald et al. (2000). 
The 6 parameters are the mass power spectrum slope
and normalization, the gas temperature at mean density $T_0$, 
the equation of state index $\gamma$, the normalization factor
$A$ (eq. [\ref{A}]), and $k_f$, which is
the smoothing scale that defines the smoothing of gas relative to dark matter.
\footnote{In principle, parameters such as $\Omega_m$ and $\Omega_\Lambda$
also affects the predicted forest properties, but at redshift $~3$,
$\Omega_\Lambda$ is negligible and 
$\Omega_m$ is close to unity. The cosmological constant
$\Omega_\Lambda$ does affect the translation between velocities
and distances -- the mass power spectrum slope and normalization
are essentially fixed at the relevant velocity scales.}
The quantity of interest for our purpose here is the combination
$A T_4^{0.7} \equiv A (T_0/10^4 K)^{0.7}$, which is directly proportional
to $(\Omega_b h^2)^2 /J_{\rm HI}$. 
From the analysis of Zaldarriaga et al. (2000),
it is straightforward to obtain the likelihood for
$A T_4^{0.7}$, marginalized over the other parameters. This is shown
in Fig. \ref{a0t0}a. It provides the starting point for our analysis below.
It is worth mentioning that the mean decrement from McDonald et al. (2000)
has not been corrected for a possible bias introduced by continuum-fitting.
However, comparing their measurement with that from
Rauch et al. (1997), who did attempt a correction (which is
model dependent), reveals that 
the bias is comparable to the error-bar given by McDonald et al.,
who gives $\langle e^{-\tau} \rangle = 0.684 \pm 0.023$ at $z=3$. 

\subsection{Likelihood for the Baryon Fraction}
\label{likelihood}

The likelihood for $A T_4^{0.7}$ shown in Fig. \ref{a0t0}a can, in principle,
be translated directly into a likelihood for $\Omega_b h^2$ after
marginalization over other parameters that appear in $A$ (eq. [\ref{A}]),
including $J_{\rm HI}$, $h$, etc. However, as mentioned in \S \ref{intro},
there are at least two reasons why, strictly speaking, one obtains only lower
bounds for the baryon fraction -- first, not all baryons are necessarily in the
forest, although a large fraction does appear to be (i.e. $\Omega_b h^2$ in
eq. [\ref{A}] refers to only those baryons in the forest); second, one often
has strictly speaking only lower limits on the ionizing background $J_{\rm
HI}$, based on summing over contributions from known sources (an exception is
the use of the proximity effect; another exception is an observational {\it
upper} limit on $J_{\rm HI}$ from searches for fluorescent Ly$\alpha$ emission;
both are discussed below).  In this section, we will discuss both the
likelihood for $\Omega_b h^2$ -- as if all or most baryons are in the forest,
and the total $J_{\rm HI}$ is known -- as well as the lower-bound-likelihood
for $\Omega_b h^2$.

\subsubsection{A Probabilistic Analysis of Lower Bounds}
\label{lowerbound}

Let us begin by developing the idea of a lower-bound-likelihood.  Previous work
often gave lower bounds on the baryon fraction as if they were strict
bounds. Of course, uncertainties in $J_{\rm HI}$, in the measured mean
decrement, and in other parameters such as $T_0$ imply that such bounds have to
be given a probabilistic interpretation.  We show here the proper quantity to
consider is {\it a lower limit on the probability that $\Omega_b h^2$ is larger
than some value}.

We have from eq. (\ref{A}) 
$\Omega_b h^2 = \sqrt{A T_4^{0.7} J_{\rm HI} f(q)}$, 
where $q$ represents the parameters
$h, \Omega_m, \Omega_\Lambda$, and $f(q)$ is some function
which can be read off from eq. (\ref{A}) (we will focus
on $z = 3$ in this paper). 
The probability that $\Omega_b h^2$ is larger than some value $B_0$ is
given by
\begin{eqnarray}
\label{B0}
&& \int_{B_0}^\infty P(\Omega_b h^2) d(\Omega_b h^2) \\ \nonumber
&& = \int dq d(A T_4^{0.7}) P (q) P (A T_4^{0.7}) 
\int_{B_0^2/{A T_4^{0.7} f(q)}}^\infty P_{\rm all} (J_{\rm HI} ) 
dJ_{\rm HI}
\\ \nonumber
&& \ge \int dq d(A T_4^{0.7}) P (q) P (A T_4^{0.7}) 
\int_{B_0^2/{A T_4^{0.7} f(q)}}^\infty P(J_{\rm HI}) dJ_{\rm HI}
\\ \nonumber 
&&
\equiv P_{>} (\Omega_b h^2 > B_0)
\end{eqnarray}
where $P(q), P(A T_4^{0.7}), P(J_{\rm HI})$ represent the
probability distributions of $q$, $A T_4^{0.7}$ and $J_{\rm HI}$.
We use $P_{\rm all} (J_{\rm HI})$ to denote the probability distribution
for the ionizing intensity from {\it all} possible sources, while the 
probability
distribution $P(J_{\rm HI})$ that we will use,
at least for some of the calculations below,
comes from a summation over contributions
from {\it known} sources, including Lyman-break galaxies and quasars.
The inequality above can be understood as follows. 
For any random variable $Z = X + Y$ with $X, Y \ge 0$ and 
independent, we have
$\int_{Z_0}^\infty P_Z (Z) dZ =
\int_0^\infty dY P_Y (Y)
\int_{Z_0-Y}^\infty dX P_X (X)
\ge \int_0^\infty dY P_Y (Y)
\int_{Z_0}^\infty dX P_X (X)
= \int_{Z_0}^\infty dX P_X (X)$.
By the same reasoning, the fact that the amount of baryons in the forest is 
necessarily smaller than the cosmological total only serves to 
strengthen the inequality in eq. (\ref{B0}).
We introduce the symbol $P_{>} (\Omega_b h^2 > B_0)$ to denote
the lower limit to the cumulative probability that $\Omega_b h^2$ is larger
than some value $B_0$. 

\begin{figure}[htb]
\centerline{\psfig{figure=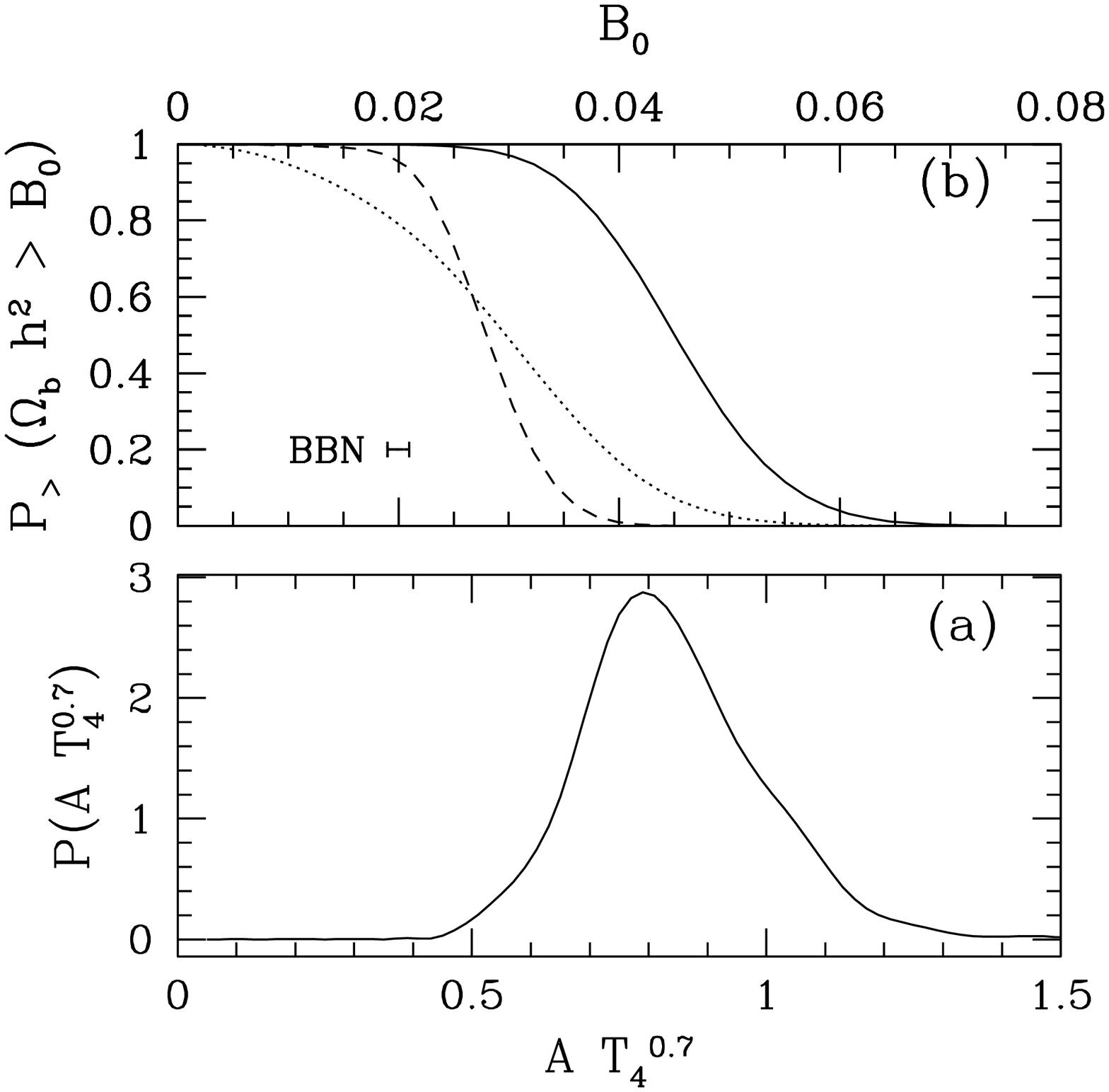,height=5.0in}}
\caption{\label{a0t0} (a) Lower panel: 
likelihood of $A {(T_0/10^4{\,\rm K})}^{0.7}$ at $z=3$.
(b) Upper panel: lower limit to the cumulative probability
that $\Omega_b h^2$ is larger than some value
$B_0$ (eq. [\ref{B0}]). 
Dotted, solid and dashed lines correspond to choices 1, 2 and 3 for
$P(J_{\rm HI})$ respectively (see text). The horizontal bar
labeled by BBN gives the nucleosynthesis $68 \%$ region.
}
\end{figure}

There are therefore, three probability distributions that we have to specify:
$P(A T_4^{0.7})$ is already given in Fig. \ref{a0t0}a -- that is the
input from observations of the Ly$\alpha$ forest;
$P(q)$ does not actually affect our conclusions very much,
but we will assume a flat universe, with independent
Gaussians for $\Omega_m h$ ($0.28 \pm 0.09$; Eisenstein \& Zaldarriaga
2001) and $h$ ($0.72 \pm 0.08$; Freedman et al. 2001)
(at $z=3$, $\Omega_\Lambda$ plays essentially no role); 
$P(J_{\rm HI})$ is what we turn to next. There are
three main sources of information.

A recent analysis of proximity effect by Scott et al. (2000)
(see also Giallongo et al. 1996, Cooke, Epsey \& Carswell 1997)
gave $\Gamma_{\rm HI} = 1.9^{+1.2}_{-1.0} \times 10^{-12} {\, \rm s^{-1}}$,
averaged from $z = 1.7$ to $3.8$, 
which translates into $J_{\rm HI} = 0.44^{+0.28}_{-0.23} {\, \rm s^{-1}}$. 
This is presumably the ionizing background from all possible sources.

Steidel et al. (2001) gave 
$j_{\nu_{\rm HI}} = 1.2 \pm 0.3 \times 10^{-21}
{\, \rm erg \, s^{-1} \, cm^{-2} \, Hz^{-1} \, sr^{-1}}$
from the Lyman-break galaxies (LBG's) 
at $z \sim 3$. \footnote{The quoted number was computed by
Steidel et al. assuming Einstein de-Sitter cosmology, but the value
is in fact strictly independent of cosmology.}
This is obtained by examining the
spectra of those LBG's that are in the bluest quartile
of observed LBG UV colors. As pointed out by 
Steidel et al. (2001), a conservative interpretation
would therefore be that no ionizing radiation escapes from the
other galaxies, in which case $j_{\nu_{\rm HI}}$ would be
a factor of 4 smaller: $j_{\nu_{\rm HI}} = 0.3 \pm 0.075 \times 10^{-21}
{\, \rm erg \, s^{-1} \, cm^{-2} \, Hz^{-1} \, sr^{-1}}$. 
There are in addition a number of uncertainties, including
uncertainty in the mean absorption by the intergalactic medium,
and the use of the luminosity function down to magnitudes that
are fainter than those used for the escape fraction measurements
(see Steidel et al. for details). 
These uncertainties are smaller than the factor
of $4$ uncertainty in interpretation.
To convert the
above into $J_{\rm HI}$, we use $\beta = 1.4$ (see eq. [\ref{JHI}]; we will
discuss the reason for this choice of $\beta$ in \S \ref{sec:spectra-lbg}).

Finally, there is the contribution from quasars. 
Rauch et al. (1997) discussed the various measurements and
calculations in some detail, and they arrived at
a conservative lower limit of $\Gamma_{\rm HI} > 7 \times 10^{-13} {\, \rm s^{-1}}$
($J_{\rm HI} > 0.16$)
at $z = 2 - 3$,
when they consider only known quasars with no extrapolation
to the faint end of the luminosity function.
We will treat this as a strict lower limit for $J_{\rm HI}$.
\footnote{In other words, the probability distribution of $J_{\rm HI}$ from
{\it all} quasars $P_{\rm all \,\, q.} (J_{\rm HI})$ is assumed
to satisfy the inequality $\int_{J^0_{\rm HI}}^\infty
P_{\rm all \,\, q.} (J_{\rm HI}) dJ_{\rm HI}
\ge \int_{J^0_{\rm HI}}^\infty P_{\rm known \,\, q.} (J_{\rm HI}) dJ_{\rm HI}$,
where $J^0_{\rm HI}$ can take any given value, and
$P_{\rm known \, \, q.} (J_{\rm HI})$ is the probability distribution
of $J_{\rm HI}$ from {\it known} quasars, and is modeled as a delta function
which peaks at $J_{\rm HI} = 0.175$. }

We will work out the lower-bound likelihood in eq. (\ref{B0}) 
for three different $P(J_{\rm HI})$'s: 
\begin{itemize}
\item 1. $J_{\rm HI}$ from the proximity effect;
\item 2. $J_{\rm HI}$ from summing the high value of
$j_{\nu_{\rm HI}}$ ($j_{\nu_{\rm HI}} = 1.2 \pm 0.3 \times 10^{-21}$) from LBG's and Rauch et al.'s value
from quasars;
\item 3. $J_{\rm HI}$ from summing the low value of
$j_{\nu_{\rm HI}}$ ($j_{\nu_{\rm HI}} = 0.3 \pm 0.075 \times 10^{-21}$) from LBG's and Rauch et al.'s value
from quasars.
\end{itemize}
The corresponding quoted error-bars are used in setting the probability
distributions for $J_{\rm HI}$. In all cases, quantities that are by definition
positive, e.g. $J_{\rm HI}$, $\Omega_m h$, etc, are constrained to be
so.

We show in Fig. \ref{a0t0}b
the result of combining the likelihoods for $A T_4^{0.7}$, $q$ and
$J_{\rm HI}$ as described above using eq. (\ref{B0}). 
To reiterate, the main observational inputs are measurements
of the mean decrement and transmission power spectrum of
the Ly$\alpha$ forest at $z=3$. 
The y-axis is the lower limit to the
likelihood that $\Omega_b h^2$ is larger than some value
$B_0$, $P_{>} (\Omega_b h^2 > B_0)$ as defined in eq. (\ref{B0}).
The current big bang nucleosynthesis value inferred from deuterium measurements
is $\Omega_b h^2 = 0.02 \pm 0.001$ (Burles, Nollett, Turner 2000).
This is ruled out at $79 \%$, $99.8 \%$ and $96 \%$ (or better) 
for the choices 1, 2 and 3 above for $P(J_{\rm HI})$.

The large difference between the curves in Fig. \ref{a0t0}b illustrates that
the dominant uncertainty by far
is that due to $J_{\rm HI}$. We emphasize, however, even assuming the
highest $J_{\rm HI}$ (model 2), the nucleosynthesis value is
not strongly excluded by Ly$\alpha$ forest measurements, although
there is an inconsistency at the $\sim 3 \sigma$ level.

\subsubsection{Further Investigations of the Baryon Likelihood}
\label{further}

\begin{figure}[htb]
\centerline{\psfig{figure=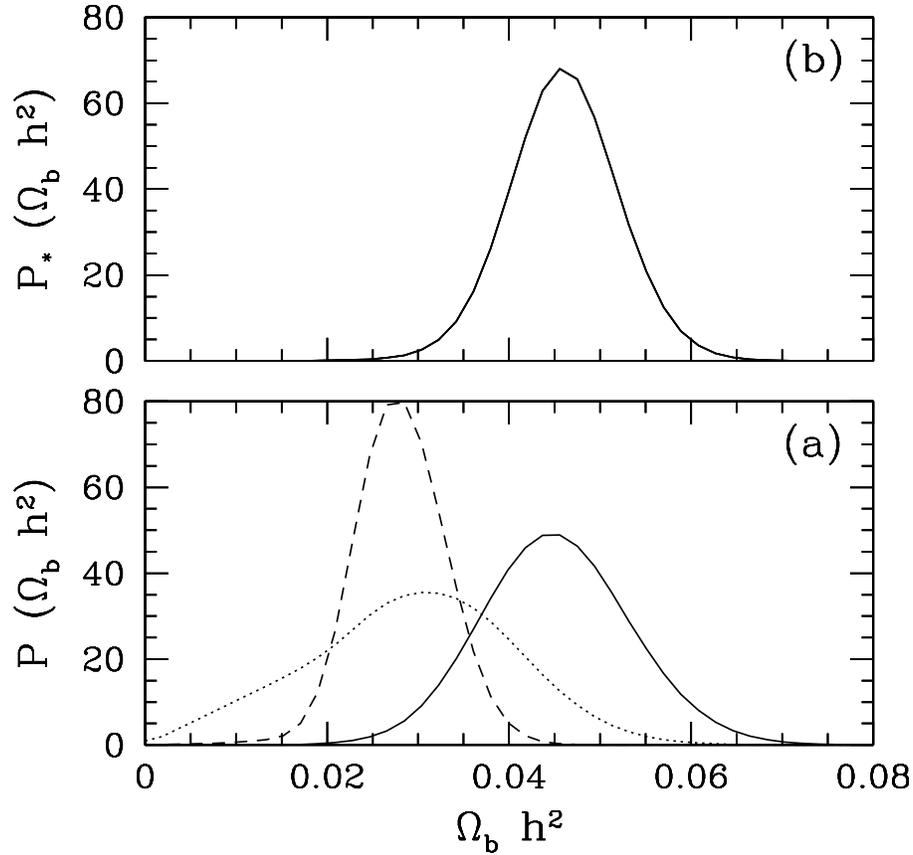,height=5.0in}}
\caption{\label{diffP}
(a) Lower panel: the likelihood (differential rather than cumulative
distribution) for $\Omega_b h^2$, where dotted, solid and dashed lines 
correspond to choices 1, 2 and 3 for $P(J_{\rm HI})$ respectively. 
These curves derive from differentiation of those in Fig. \ref{a0t0}b;
see text for details. (b) Upper panel: 
$P_* (\Omega_b h^2)$, the likelihood for
$\Omega_b h^2$ if $J_{\rm HI}$ is fixed to be $1$; this is useful
for generating the likelihood of $\Omega_b h^2$ for arbitrary
distributions of $J_{\rm HI}$ (see eq. [\ref{PstarJ}]).
$P_* (\Omega_b h^2)$ is well approximated by a Gaussian
of mean $= 0.046$ and dispersion $= 0.0061$.
For $J_{\rm HI}$ fixed to be some value different from unity, the probability
distribution for $\Omega_b h^2$ can be obtained by simply
multiplying both the mean and dispersion by $\sqrt{J_{\rm HI}}$.
}
\end{figure}

It is instructive to show a different version of Fig. \ref{a0t0}b, namely
the derivatives of those same curves (with a negative sign) i.e.
$-d P_{>} (\Omega_b h^2 > B_0) / dB_0$. 
This is done in Fig. \ref{diffP}a, whose y-axis now 
represents the differential
probability distributions for $\Omega_b h^2$ ($P(\Omega_b h^2)$, rather
than $P_> (\Omega_b h^2 > B_0)$; see eq. [\ref{B0}]),
{\it if} one assumes that the majority of the baryons is in the forest, and
that the probability distributions for $J_{\rm HI}$ used above actually
approximate the true distributions for the {\it total} $J_{\rm HI}$.
Phrased in this way, the mean values 
with dispersions for
$\Omega_b h^2$ are $0.03 \pm 0.01$, 
$0.045 \pm 0.008$ and $0.028 \pm 0.005$
respectively for the choices 1,2 and 3 for $P(J_{\rm HI})$ laid out previously.

The above represents another way of saying that the present measurements
have a weak preference for $\Omega_b h^2$ somewhat larger than the
nucleosynthesis value.
The error-bars at the moment are dominated by uncertainties in $J_{\rm HI}$. 
However, it would be useful to derive a likelihood for $\Omega_b h^2$ assuming
$J_{\rm HI}$ is exactly known. Future improvements in measurements of
$J_{\rm HI}$ can then be directly translated into improved distributions
for $\Omega_b h^2$. To be concrete, we would like to compute 
$P_* (B_0')$:
\begin{eqnarray}
\label{Pstar}
P_* (B_0') \equiv 
P (\Omega_b h^2=B_0', J_{\rm HI}=1) = 
\int dq {2 B_0' \over f(q)}P(q) P(AT_4^{0.7} = {B_0'}^2/f(q))
\end{eqnarray}
which is obtained from eq. (\ref{B0}) by setting 
$P(J_{\rm HI}) = \delta(J_{\rm HI} - 1)$.
We show $P_* (B_0'=\Omega_b h^2)$ in Fig. \ref{diffP}b. 
It is well-approximated by a Gaussian with a mean of
$0.046$ and a dispersion of $0.0061$. 
This is
the likelihood for $\Omega_b h^2$ if $J_{\rm HI}$ were exactly unity.
If $J_{\rm HI}$ were $0.5$ for instance, then the likelihood
for $\Omega_b h^2$ would be a Gaussian with a mean of
$\sqrt{0.5} \times 0.046 = 0.033$ and a dispersion of 
$\sqrt{0.5} \times 0.0061 = 0.0043$. 
More generally,
given any $P(J_{\rm HI})$, the differential 
probability distribution for $\Omega_b h^2$ can be easily computed
using $P_*$:
\begin{equation}
P(\Omega_b h^2) = \int {1 \over \sqrt{J_{\rm HI}}} P_* 
(B_0' = \Omega_b h^2/\sqrt{J_{\rm HI}}) 
P(J_{\rm HI}) d J_{\rm HI}
\label{PstarJ}
\end{equation} 
Or, more rigorously, in cases where $P(J_{\rm HI})$ accounts for
only contributions from known but not all possible sources,
and keeping in mind that not all baryons are
necessarily in the forest, one can obtain (by integrating
eq. [\ref{PstarJ}]) a lower bound on the cumulative
probability that $\Omega_b h^2$ is larger than some value -- i.e. use
$\int_{B_0}^\infty P(\Omega_b h^2) d(\Omega_b h^2)$ to obtain
$P_> (\Omega_b h^2 > B_0)$, 
just as we have done in 
\S \ref{lowerbound}. 

Fig. \ref{diffP}b illustrates another important point: 
substantial uncertainties in $\Omega_b h^2$ remain 
even if $J_{\rm HI}$ were known at high accuracy. 
The uncertainties can be traced back to the spread
in $A T_4^{0.7}$, shown in Fig. \ref{a0t0}a. 
Checking each of the parameters that were marginalized
in obtaining $P(A T_4^{0.7})$, we find that the main
sources of errors are, in order of importance,
$T_0$ the temperature at mean density, 
$\gamma$ the equation of state index, and $k_f$ the
smoothing scale of the gas distribution --
fixing each of these parameters individually would decrease
the variance in $A T_4^{0.7}$ by $43 \%$, $39 \%$
and $31 \%$ respectively. Interestingly,
fixing the power spectrum normalization and/or slope,
or the mean transmission value ($1 -$ mean decrement) decreases
the variance of $A T_4^{0.7}$ by no more than $11 \%$. 
The latter in particular implies that the current
accuracy of the mean transmission/decrement measurements
at $z \sim 3$ is not the main limiting factor for the 
the accuracy of the $\Omega_b h^2$ constraint; one caveat is that
systematic errors such as those due to continuum-fitting might
have been underestimated -- this is an important question
which deserves closer study.
In addition to better measurements
of $J_{\rm HI}$, future improvements in constraints
on $\Omega_b h^2$ from the Ly$\alpha$ forest will rely
on more accurate knowledge of the thermal state
of the IGM.

\section{The Ionizing Continuum of Lyman Break Galaxies}
\label{LBO}

The most stringent lower limit on $\Omega_b h^2$ (dashed curve
in Fig. \ref{a0t0}a) depends
sensitively on the ionizing continuum ($E>13.6$eV) emitted by 
high--redshift galaxies.  This is obtained
from the observed composite spectrum of 29 Lyman break galaxies 
at $z \sim 3.4$ 
(Steidel et al. 2001).
\footnote{The Steidel et al. measurements strictly speaking
constrain $J_{\rm HI}$ at $z \sim 3.4$ rather than $3$, as we
have used in \S \ref{likelihood}. Within the present
uncertainties, $J_{\rm HI}$ does not appear to evolve
significantly with redshift around $z \sim 3$ 
(see e.g. Giallongo et al. 1996, Cooke et al. 1997, Scott et al. 2000). 
We choose to focus on $z = 3$ partly because
some of the best existing measurements on mean decrement
and transmission power spectrum are at $z = 3$.}
Because of the importance of
the ionizing continuum in determining the likelihood for $\Omega_b
h^2$, we discuss here the results of Steidel et al. (2001) in some
more detail.  In particular, the continuum flux in the Steidel et
al. (2001) sample is unexpectedly high (compared
with both naive theoretical expectations, and
observations of galaxies at low redshifts; see below), 
and, taken at face value,
presents two distinct puzzles: (i) a significant fraction
of the ionizing radiation appears to escape from the LBGs, 
whereas a small escape fraction is expected; and (ii) even 
assuming that no ionizing photons were lost in
escaping from the galaxies, the stellar populations of the LBGs appear
to produce a very hard ionizing continuum compared to spectral
synthesis models.  We address these two issues below, first (ii),
and then (i).

\subsection{The Ionizing Flux Produced in Lyman Break Galaxies}
\label{sec:spectra-lbg}

We first compare the stacked composite spectrum of LBGs
in Steidel et al. (2001) with model
galactic spectra. We utilize the spectral synthesis models of
Leitherer et al. (1999); although we obtained similar conclusions
using the models of Bruzual \& Charlot (1996).  We adopt the stellar
models for a Salpeter IMF (with a slope $\alpha=2.35$) between
$1-100$M$_\odot$, metallicity of $Z=0.4$Z$_\odot$, and compute the
spectrum either for continuous star formation for $10^8$ yr, or at the age of
$10^6$ yr following an instantaneous starburst.  These correspond to
the models shown in Figures 8d and 7d of Leitherer et al. (1999),
respectively.  In addition, to simulate the observed composite
spectrum, we modify the emitted stellar template
spectrum by including the effects of dust absorption, as well as 
the opacity of the
intervening IGM. To include dust opacity, we use the absorption           
cross--sections for a mix of graphite and silicate grains that
reproduces the Milky Way opacities (Draine \& Lee 1984), and adjust
the total amount of dust by fitting the slope of the observed spectrum
at (emitted) wavelengths $\lambda>1215$\AA.

In order to assess the intrinsic level of the flux below 912\AA, it is
important to model the opacity of the intervening IGM accurately. This opacity
is due to Lyman $\alpha$ absorption systems along the lines of sights to the
galaxies.  We model here the IGM absorption by summing the optical depths of
all Lyman $\alpha$ absorbers in the range of neutral HI column densities
$10^{12}-10^{20}$ cm$^{-2}$ (see Madau 1995 for a description of the method,
and Fardal, Giroux \& Shull 1998 for a more recent parameterization of the
column density distribution of absorbers).  Note that the opacity at
wavelengths shorter than 912\AA\, is dominated by the poorly--known abundance
of high--column density absorbers ($N_{\rm HI}\gsim 10^{17}~{\rm cm^{-2}}$).
However, the relevant integrated quantity that determines the break in the
spectrum at 912\AA\, is better constrained.  Using a quasar sample, Steidel et
al. (2001) measured the value of the intrinsic ratio of the decrements at
900\AA\, and 1100\AA, which they find to be a factor of 2.5.  In order to match
this decrement ratio, and to simultaneously predict a Ly$\alpha$ decrement of
$D_A\approx 0.45$ (which we obtained directly from the $z=3.4$ LBG sample), we
found that we had to adopt a column--density distribution intermediate between
models A1 and A4 of Fardal et al. (1998).

In Fig.~\ref{fig:spectra}, we show the resulting model spectra,
superimposed on the composite spectrum of Steidel et al. (2001).  The
dotted and dashed curves correspond to the models of continuous star
formation for $10^8$ years, and to the $10^6$ yr old starburst,
respectively; the solid curves show the data.  All fluxes are
normalized to the same flux at the emitted wavelength of 1500\AA. In the
bottom panel, we also show the optical depths we assumed for the dust
and the IGM, respectively (in the continuous star formation case; a
somewhat larger dust opacity was needed for the instantaneous
starburst).  In the top panel, the horizontal dashed line shows the
mean observed flux in the wavelength range $880\AA\leq \lambda\leq
910\AA$.

As the top panel in Fig.~\ref{fig:spectra} shows, in the $880\AA\leq
\lambda\leq 910\AA$ range, the mean observed flux is
$F_\nu/F_{1500}=0.06\pm 0.01$; while the continuous star formation
models predict $F_\nu/F_{1500}=0.027\pm 0.003$. Taken at face value,
the data reveals twice as high an ionizing flux as these models
predict, even if we do not include any absorption by cold gas in the
galaxy.  The starburst model can come close to producing the observed
ionizing flux at 912\AA, but only if an age of $\lsim 10^6$yr is
assumed.  Although the 29 galaxies included in the composite spectrum
were selected from the bluest quartile of the sample, it appears
unlikely that they were typically ``caught'' at such a young age. A
short lifetime does appear to be consistent with the number density
and clustering properties of LBG's (e.g. Wechsler et al. 2001; but see
also 
Mo, Mao \& White 1999), but it
is disfavored by other arguments, such as the optical to
near-infrared colors (Shapley et al., in preparation), and by
tentative measurements of mass--to--light ratios (Pettini et
al. 2001). The ionizing radiation in
the models is dominated by massive OB stars, and, as a result, after
$\sim 4\times10^6$ years, the starburst model predicts a rapidly
diminishing value for the flux below 912\AA.

The starburst model gives a (cross-section weighted) 
spectral slope of $\beta = 1.4$ below
$912 \AA$ ($j_\nu \propto \nu^{-\beta}$). 
This slope is somewhat harder than typical quasar spectra in
the relevant wavelength range just below $912 \AA$.
This is what we have
used in \S \ref{likelihood} to convert measurements of $j_{\nu_{\rm HI}}$ to
$J_{\rm HI}$. The true value for $\beta$ is likely to be smaller 
due to absorption by the intergalactic medium
-- this gives
us a lower limit on $J_{\rm HI}$ ($J_{\rm HI} \propto 1
/ (\beta + 3)$; see eq. [\ref{JHI}]). 

Taking the data at face value, the implication is that the stellar
populations in LBGs produce $\gsim$twice as many H--ionizing photons
as the current models predict.  
This discrepancy is more severe at shorter wavelengths, where the IGM
is expected to become increasingly more opaque.  Note that although the mean
IGM opacity is directly measured by Steidel et al. at $\lambda\approx 900\AA$,
the variation of the opacity with wavelength across the range $880-910\AA$
depends on the column density distribution of Lyman $\alpha$ absorption systems
with column densities $N_{\rm HI}\gsim 10^{17}~{\rm cm^{-2}}$, and is more
uncertain.                                  

Changing the assumed metallicities, or
adding the emission from the nebular continuum, does not significantly
increase the predicted ionizing fluxes.  The UV fluxes of OB stars can
be increased by the presence of stellar winds (not included in the
stellar models we adopted).  Although winds can increase the
HeII--ionizing flux by orders of magnitude, the corresponding increase
for the H--ionizing flux in O stars has been found to be small
(Schaerer \& de Koter 1997). The increase can be more significant for
cooler B--stars (Najarro et al. 1996), but these stars do
not dominate the ionizing photon budget in the continuous star--formation
models, in which O--stars are continuously replenished.  Furthermore,
the hydrostatic stellar models typically reproduce the properties of
Galactic HII regions (Leitherer et al. 1999), so that an increase by
the required factor of $\sim 2$ would make it more difficult to
reconcile the models with these observations.

In summary, we conclude that the stellar population in the galaxies
comprising the Steidel et al. (2001) sample appear to be significantly
different from spectral synthesis model predictions, and from stellar
populations in nearby galaxies, unless the selection procedure
somehow strongly favors very young galaxies (i.e. age $\lsim 10^6$ years).
It remains to be seen whether stellar
models that include winds can account for the apparent difference.  An
alternative explanation of the large ionizing flux could be that the
IMF at $z \sim 3$ is biased towards massive stars.

\subsection{The Escape Fraction of Ionizing Photons From Lyman Break Galaxies}
\label{sec:fesc-lbg}

Although the nature of LBGs is not well understood, based on their
observed fluxes and clustering properties, they appear to have
luminosities similar to $\sim$L$^*$ galaxies, and to be associated with
halos of total mass $\gsim 10^{11}$M$_\odot$ (in most models, see,
e.g. Wechsler et al. 2001, and references therein).  
The escape of ionizing radiation from
``normal'' disk galaxies has been studied extensively in
observations of the Milky Way (Reynolds et al. 1995) and other nearby
galaxies (e.g. Leitherer et al. 1995; Hoopes, Walterbos \& Rand 1999);
as well as theoretically (e.g. Dove \& Shull 1994; Dove, Shull \&
Ferrara 2000).  Both theory and observations indicate small escape
fractions that are of order $\sim 10\%$.  In a theoretical study
extending galactic disk models to high redshift, Wood \& Loeb (2000)
showed that if a few percent of the total gas mass would cool and
settle into a rotationally supported thin disk, the escape fraction
would be negligible from galaxies at $z\sim 3$ in halos with a total
mass $M\gsim 10^{11}$M$_\odot$ (the decrease in $f_{\rm esc}$ being
due to the higher densities [$\propto(1+z)^3$] at higher redshifts).
In comparison, the escape fraction in the Steidel et al. sample
is $f_{\rm esc}\gsim 0.5$.
\footnote{Steidel et al. define $f_{\rm esc}$ as
the fraction of 900\AA\, photons that escape the galaxy, divided by 
the same fraction for 1500\AA\, photons.  Thus, the actual escape
fraction of 900\AA\, photons from the LBGs can be as small as $\sim 15\%$, 
if dust absorbs $\sim 85\%$ of photons at both wavelengths.  Since
dust opacity is not included in the theoretical models,
this is a fair comparison.}

Ionizing radiation could escape more efficiently from LBGs if most of
the cold gas ($T \sim 10^4$ K) 
were spread over an extended region, rather than
incorporated into a fully--formed, dense galactic disk.  Such a
spatially extended distribution for the cold gas would be natural
during the earliest phases of galaxy formation, when galaxy--sized
dark halos are assembling for the first time.  Alternatively, if star
formation in LBGs is triggered by mergers (see this model in Wechsler
et al. 2000), then the merger could spread the cold gas over a large
solid angle.  In either case, the diluted density would reduce the
recombination rate ($\propto\rho^2$), making it easier to keep the
cold gas photoionized, and for the ionizing radiation to escape.

To quantify the required ionizing photon rate, we consider the
simplified toy model for the cold gas distribution described in Haiman \&
Rees (2001).  In this model, the gas initially has a spherically
symmetric distribution, with a radial profile dictated by hydrostatic
equilibrium in an NFW (Navarro, Frenk \& White 1997) dark matter halo.
Subsequently, a fraction of the gas, determined by the cooling
efficiency, cools to $10^4$K, and resides in cold clumps whose
densities are compressed by a factor $\sim T_{\rm vir}/10^4$K (where
$T_{\rm vir}\sim10^6$K is the virial temperature of the halo).  The
rest of the gas remains in the collisionally ionized hot phase, at the
temperature of $T_{\rm vir}$.

As an example of this model, the gas in a $10^{12}$M$_\odot$ halo at
$z=3.4$ (virial temperature of $T_{\rm vir}=2\times10^6$K, a
$2.3\sigma$ object) cools out to approximately half of the virial
radius ($R_{\rm vir}\approx50$kpc).  The total recombination rate
within the halo is $R=4\pi\int_0^{R_{\rm vir}} r^2 dr f_V n_{\rm H}(r)^2
\alpha_B$ is $\approx 2\times10^{55}$s$^{-1}$, where $f_V$ is the
volume filling fraction of cold gas (approximately unity
out to $R_{\rm vir}/2$, and dropping to negligible values
at larger radii), and $\alpha_B$ is the case-B
hydrogen recombination coefficient evaluated at $\approx 10^4$K.
With the standard Salpeter IMF, every stellar proton yields $\sim$
4000 ionizing photons, so that a star formation rate of 1 M$_\odot$
yr$^{-1}$ translates to a production rate $1.5\times10^{53}$ s$^{-1}$
of ionizing photons.  
In order to allow a large escape fraction, most of this cold gas has
to be kept photo--ionized.  Increasing the above photon production
rate by a factor of four, as suggested by the composite spectrum of
Steidel et al. (2001), we find that the required ``characteristic''
star formation rate is $\sim 40$M$_\odot$ yr$^{-1}$.
This star formation rate is consistent with the luminosities of
LBGs. Furthermore, this rate converts only $\sim 2\%$ of the available gas
into stars in $10^8$ yr, making our assumption of continuous star
formation for $\sim 10^8$ years plausible.  We also note that in this
model, the required star formation rate for a halo of mass $M$ at
redshift $z$ scales approximately as
\begin{equation}
\dot M_{\star}\approx 40 \frac{\rm M_\odot}{\rm yr}                  
\left(\frac{M_{\rm halo}}{10^{12}{\rm M_\odot}}\right)^{5/3}
\left(\frac{1+z}{4.4}\right)^{4},
\label{eq:sfr}
\end{equation}
making it easier for ionizing radiation to escape from smaller halos
(see Haiman \& Rees 2001 for details).

The model described above is highly idealized, and should be
regarded only as a plausibility argument that large escape fractions are
possible with an extended gas distribution.  These models predict that
the photoionized cold gas could be detectable in the Ly$\alpha$
recombination line, as a low surface brightness ``fuzz'' with $\gsim
10^{-18}~{\rm erg~s^{-1}~cm^{-2}~asec^{-2}}$. Deep Ly$\alpha$ imaging
of extended regions around the LBGs can therefore furnish a method to
test this model (see Haiman \& Rees 2001 for a similar discussion in
the context of bright quasars).  

An alternative scenario is one in which the cold gas
resides in clumps with a small covering factor. This could allow most
of the ionizing photons to escape along lines of sights traversing
only hot, collisionally ionized, medium.  However, in a 
self--gravitating two--phase medium, this explanation
requires a minimum cold clump size that exceeds the Jeans mass of the
cold phase (Rees 1988). 
As a result, the postulated large clumps would be unstable to
fragmentation, which would tend to increase the covering factor to
unity.  Nevertheless, the covering factor of cold gas can be decreased by
feedback mechanisms. Indeed, Pettini et al. (2001) found evidence for
galactic scale outflows (``superwinds'' of several hundred km/s) in
the spectra of LBGs to be ubiquitous. As noted in Pettini et
al. (2001), such superwinds, driven by mechanical energy from
supernova and stellar winds, can punch holes through the galactic
disk, through which the ionizing radiation can escape.  

\section{Discussion}
\label{discuss}

To summarize, the accuracy of the present constraints on the baryon fraction
from the Ly$\alpha$ forest is limited by our knowledge of the ionizing
background. A slight tension does exist between such constraints and
the nucleosynthesis value, in that the forest measurements tend
to prefer higher values for $\Omega_b h^2$. However, even assuming
the largest estimate of the ionizing background, 
which is dominated by contributions from Lyman-break galaxies as reported by
Steidel et al. (2001) (see model 3 in \S \ref{lowerbound}), 
the nucleosynthesis value is ruled out only at 
the $3 \sigma$ level. 

We emphasize also that even if the ionizing background
is known to high accuracy, substantial uncertainties remain, largely due
to the uncertain thermal state of the intergalactic medium (parametrized by $T_0$ and
$\gamma$ -- the temperature at mean density and the equation
of state index; see \S \ref{model}).
Present measurements give a probability distribution of
$\Omega_b h^2$ which is well-approximated by a Gaussian:
\begin{equation}
P_* (\Omega_b h^2) = {1\over \sqrt{2 \pi} 0.0061} {\rm \,exp \,} 
[-(\Omega_b h^2 - 0.046)^2/0.0061^2]
\label{PstarOK}
\end{equation}
{\it if} $J_{\rm HI}$ is fixed to be unity. We show in eq. (\ref{PstarJ})
how the above probability can be used to obtain the probability
distribution for $\Omega_b h^2$, $P(\Omega_b h^2)$, given any distribution for
$J_{\rm HI}$ (in the extreme case, a delta function).
This will be useful as measurements of $J_{\rm HI}$ improve.
We also show in \S \ref{lowerbound} how the lower bound likelihood
can be obtained from a given $P(\Omega_b h^2)$ -- this is
a more appropriate quantity to consider if only strictly
lower limits to the ionizing background are known.

Our results show that future improvements in the Ly$\alpha$ forest constraint
on the baryon fraction will have to rely on improvements in measurements of the
ionizing background and of the thermal state of the gas.  The former can be
achieved through a variety of means.  Measurements of the ionizing flux from
LBG's extending to the redder population and fainter magnitudes will be useful.
Improved measurement of the proximity effect would, in principle, be
especially 
powerful because it measures the sum total background of all sources.
However, one main source of uncertainty is the large velocity shifts between
different lines which affects estimates of the effect as a function of distance
from quasars (see e.g.  Scott et al. 2000). In addition, 
systematic effects due
to clustering and peculiar velocities 
should be taken into account in the 
translation of absorption line statistics into the background flux 
-- interestingly, Loeb \& Eisenstein (1995) argued that these effects
generally lead to an {\it over}-estimate
of $J_{\rm HI}$ from standard proximity effect 
measurements.
It is also useful to keep in mind the possibility of measuring $J_{\rm
HI}$ from direct imaging of the fluorescent Ly$\alpha$ photons from optically
thick systems (Gould \& Weinberg 1996). The measurements of Bunker, Marleau \&
Graham (1998) have already put an {\it upper} limit on the ionizing background,
which at the moment is still quite large (3-sigma upper limit is $j_{\rm HI} <
2 \times 10^{-21} {\,\rm erg \, s^{-1} \, cm^{-2} \, Hz^{-1} \,
sr^{-1}}$). Deeper exposure to seek a definite detection would be very useful.
As far as the thermal state of the intergalactic medium is concerned, larger
sample size is crucial, and several techniques have proven useful for this
purpose (e.g. Ricotti, Gnedin \& Shull 2000, Bryan et al. 1999, Schaye et
al. 1999, McDonald et al. 2001, Choudhury et al. 2000, Zaldarriaga et al. 2000;
but see also Meiksin, Bryan \& Machacek 2001).  We should also mention
that a constraint on the
baryon density can
be obtained from the He II forest as well
(e.g. Wadsley, Hogan \& Anderson 2000).

Ultimately, as the error-bars on $\Omega_b h^2$ from the forest come
down, we will have a very precise test of the gravitational instability
model, for the forest in particular 
as well as for large scale structure in general. 
A statistically significant disagreement among constraints from the forest,
nucleosynthesis and the microwave background would require some basic
revision of our model.

The findings of Steidel et al. (2001) regarding the LBG Lyman-continuum deserve
further study.  Their measurements imply an intrinsically hard stellar
spectrum, which seems to require a (perhaps unreasonably) young population of
stars (1 million years or so) and/or an initial mass function that is biased
towards massive stars. Moreover, the large escape fraction for ionizing photons
imply that the cold, optically thick gas in LBG's is likely distributed quite
differently from what has been seen in local galaxies. Possibilities include a
model in which the cold gas is spatially extended over the halo, and has not
settled into a dense galactic disk; or a disk with large holes punched out by
strong stellar winds, as suggested by Pettini et al. (2001).  

Finally, it is interesting to note that there are some indications of
a similar evolutionary trend in AGN -- a decrease of the escape
fraction at lower redshifts. Searches for intrinsic absorption at the
Lyman-edge at intermediate redshifts ($0.4 \lsim z \lsim 2$) 
generally turn up only small Lyman edge
discontinuities in a small fraction of quasars (see e.g. Koratkar,
Kinney \& Bohlin 1992, Zheng et al. 1997). On the other hand, there is
mounting evidence of strong internal Lyman edge absorption in nearby
Seyfert galaxies (Alexander et al. 1999, 2000; Kraemer et al. 1999).

\acknowledgements

\begin{figure}[htb]
\centerline{\psfig{figure=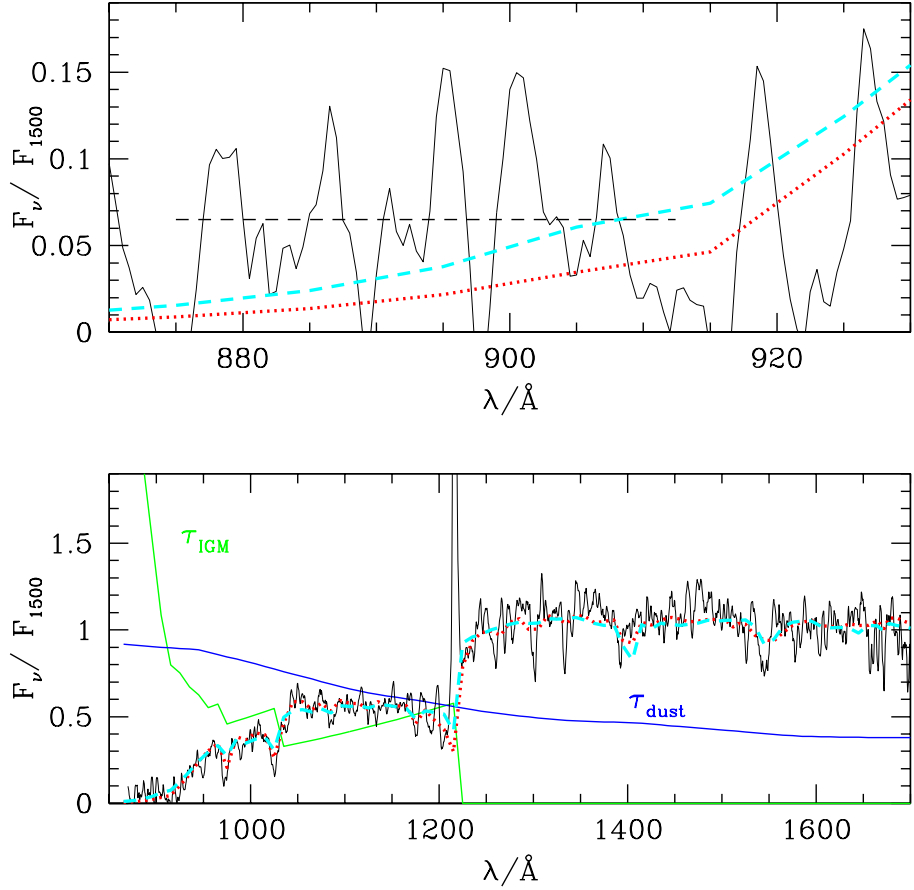,height=5.0in}}
\caption{\label{fig:spectra}
Population synthesis spectra, including opacities of dust and neutral
hydrogen in the IGM, compared with the observed spectrum of Steidel et
al. (2001).  The dotted and dashed curves correspond to models of
continuous star formation for $10^8$ years, and to an instantaneous
starburst at $10^6$ yr, respectively. The solid curves show the
Steidel et al. (2001) composite spectrum.  All fluxes are normalized
to the flux at the emitted wavelength of 1500\AA.  For reference, in
the bottom panel, we show the optical depths assumed for the dust and
the IGM.  In the top panel, the horizontal dashed line shows the mean
observed flux in the interval $880\AA\leq \lambda\leq 910\AA$.}
\end{figure}

We thank C. Leitherer and M. Pettini 
for useful discussions, and C. Steidel for electronic
data for the observational spectrum in Fig.~\ref{fig:spectra}.  LH was
supported by an Outstanding Junior Investigator Award from DOE. ZH was
supported by NASA through the Hubble Fellowship grant HF-01119.01-99A, awarded
by the Space Telescope Science Institute, which is operated by the Association
of Universities for Research in Astronomy, Inc., for NASA under contract NAS
5-26555.

\newpage


\end{document}